\documentclass[pre,a4paper,superscriptaddress,twocolumn,showpacs,amsmath,amssymb,floatfix]{revtex4}

\usepackage{graphicx}
\usepackage{amssymb} 
\usepackage{float}
\usepackage{hyperref}

\usepackage{color}
\newcommand*{\defeq}{\stackrel{\text{def}}{=}}

\begin{document}

\title{Kicked fluxonium with quantum strange attractor}

\author{Alexei D. Chepelianskii}
\affiliation{\mbox{LPS, Universit\'e Paris-Sud, CNRS, UMR 8502, Orsay F-91405, France}}
\author{Dima L. Shepelyansky}
\affiliation{\mbox{Univ Toulouse, CNRS, Laboratoire de Physique Th\'eorique,
    Toulouse, France}}

\date{December 21, 2025}

\begin{abstract}
The quantum dissipative time evolution
of a fluxonium  under a pulsed field (kicks)
is studied numerically and analytically.
In the  classical limit the system dynamics
is converged to a strange chaotic attractor.
The quantum properties of this system are studied for the density
matrix in the frame of Lindblad equation.
In the case of dissipative quantum evolution
the steady-state density matrix
is converged to a quantum strange
attractor being similar to the classical one.
It is shown that depending on the dissipation
strength there is a regime when the eigenstates
of density matrix are localized
at a strong or moderate dissipation. At a weak dissipation
the eigenstates are argued to be delocalized
being linked to the Ehrenfest explosion of quantum wave packet.
This phenomenon is related with the Lyapunov exponent
and Ehrenfest time for the quantum strange attractor.
Possible experimental realisations of this quantum
strange attractor with fluxonium  are discussed.
\end{abstract}

%

\maketitle

\section{Introduction} 
\label{sec1}

The fluxonium was invented in \cite{devoret2009}
as a single Cooper-pair circuit free of charge offsets.
Recently very long coherence times and extremely high fidelity
have been realized with fluxonium qubits
(see e.g. \cite{lyon,nguyen,alibaba,quentin2023,vavilov2024,gatemonium}).
Roadmap for development of  high-performance fluxonium quantum processor
is advanced in \cite{blueprint}.
Thus the progress with superconducting fluxoniom systems
allows to perform outstanding control of these quantum  circuits.

The Hamiltonian of fluxonium, written in the standard notations \cite{devoret2009}, reads:
\begin{equation}
  \hat{H} =  4 E_C \hat{N}^2 + E_L \hat{\varphi}^2/2 - E_J \cos(\hat{\varphi} - 2\pi \Phi_{ext}/\Phi_0)
\label{eq1}
\end{equation}
where the reduced charge on the junction capacitance is described by $\hat{N} = \hat{Q}/2e$
with conjugated flux $\varphi= 2e\hat{\Phi}/\hbar$
and charge energy $E_C$
(in units of 2$e$), the Josephson energy is $E_J$, 
shunted by a large inductance $L$, $\Phi_{ext}$ is external flux
and $\Phi_0$ is a flux quantum. There is a usual operator commutator relation
$[\hat{\varphi},\hat{N}]=i$. Typical experimental parameters
are $E_L \sim 0.5$ GHz, $E_J \sim 9$ GHz and $E_C \sim 2.5$ GHz \cite{devoret2009}
with  their certain  variation in other experiments
\cite{lyon,nguyen,alibaba,quentin2023,vavilov2024,gatemonium}.
A general introduction to physics of superconducting qubits
can be find in \cite{shnirman,wendin}.

In this work we introduce and study the kicked fluxonium model
described by the time dependent Hamiltonian
\begin{equation}
  \hat{H} =  4 E_C \hat{N}^2 + E_L \hat{\varphi}^2/2 - J \cos(\hat{\varphi} - 2\pi \Phi_{ext}/\Phi_0)
  \sum_m \delta(t - mT)
\label{eq2}
\end{equation}
where $\sum_m (t - mT)$ is a train of periodic $\delta$-functions following
with a period $T$ and producing kicks of fluxonium,
$J=E_J \delta t$ is the kick amplitude which in a case of pulse 
is determined by a finite pulse duration $\delta t$. Between kicks
the time evolution is described by a quantum oscillator
with frequency $\Omega = 2\sqrt{2 E_C E_L}$ and $\hbar=1$.
Thus the system represents a kicked harmonic oscillator which in dimensionless
units is described by the rescaled fluxonium Hamiltonian
\begin{equation}
  \hat{H_f} = (\hat{p}^2+\hat{x}^2)/2  - K \cos(q \hat{x})\sum_m \delta(t - m T) \; .
\label{eq3}
\end{equation}
Here $K/\hbar$ describes the number of kick quanta
excited by a kick from the oscillator ground state.
At $q=1$ in physical units of Hamiltonian (\ref{eq2})
we have $K/\hbar= J/\Omega$ since $\hbar=1$ in (\ref{eq2}).
In absence of kicks at $K=0$ the Hamiltonian $\hat{H_f}$ is reduced to the standard Hamiltonian of
quantum harmonic oscillator $\hat{H_0} =  (\hat{p}^2+\hat{x}^2)/2$, $[\hat{p},\hat{x}]=-i\hbar$.
The mass and frequency of oscillator are normalized to unity
so that in these units $\hbar$ is dimensionless and $q$ is also  dimensionless.
A transition between the case with $q$ to a case with $q=1$ is given by
the transformation:
$qx,qp \rightarrow x,p$, $\hbar_{eff} \rightarrow \hbar q^2$,
$K/\hbar  \rightarrow K/\hbar_{eff} = K/(\hbar q^2)$.
In a classical system the case at $q \neq 1$
can be transferred to the case at $q=1$
by the transformation
$Kq^2 \rightarrow K_{cl}$, $qx, qp \rightarrow x,p$.

In fact the system of classical kicked harmonic oscillator (\ref{eq3})
had been introduced and studies in \cite{zaslav1,zaslav2}.
It is also know as Zaslavsky web map \cite{zaslavweb}.
The Hamiltonian dynamics (\ref{eq3}) depends only on two dimensionless 
parameters: classical chaos parameter $K$, that determines the kick strength
leading to hard chaos at high values, and ratio
of the period of kicks to oscillator period $2\pi$ 
being $T/2\pi = 1/R$ (we take here $q=1$). For $R=3,4,6$ the separatrix web
covers the whole phase space plane $(x,p)$,
that corresponds to a known geometric result of covering
plane by triangles, squares and hexagons.
For these $R$ values  even at very small $K$ 
there is a chaotic separatrix layer of width proportional to $K$ \cite{zaslav1,zaslav2,zaslavweb}.
For other integer $R$ values the separatrix web cannot cover
the whole plane without gaps and the properties of
chaotic layers  at small $K$ are more complex.
In contrast for high $K$, e.g. $K=7$,
there is a formation of hard chaos without
visible stability islands with a diffusive
energy growth $E = <(p^2+x^2)/2> \sim q^2 K^2 t/2$
where time $t$ is measured in number of kicks.
The generic properties of dynamical Hamiltonian chaos
are described in \cite{chirikov1979,chaosbook,ott}.

The quantum studies of Hamiltonian (\ref{eq3}) 
were reported by different groups
(see e.g. \cite{sire,dana,zoller,bambihu,kells,buchleitner,gardiner,india2024} at $q=1$).
For $M=4$ the quantum dynamics (\ref{eq3}) is reduced to the kicked Harper model
studied in \cite{lima,artuso} as discussed in \cite{sire}
and there is no quantum dynamical localization.
This is drastically different from the case of
the kicked rotator model obtained
from the quantization of the Chirikov standard map
where the classical diffusion is suppressed by
quantum interference effects leading to the dynamical localization
similar to the Anderson localization in disordered solids
(see e.g. \cite{chirikov1988,stmapschol,fishman} and Refs. therein
for the studies of kicked rotator model).
This kicked rotator  model had been realized with cold atoms in
a kicked optical lattice \cite{raizen} where
kicks were realized by short pulses of finite duration.

For the numerical studies of quantum kicked harmonic oscillator
a number of interesting results have being obtained
with localization and delocalization
at small and high kick strength values of $K/\hbar$ 
for $R=5$ and irrational $R$ \cite{sire,kells}.
The experimental realization of quantum system (\ref{eq3})
with ion trap was proposed in \cite{zoller}
with analysis of sensitivity and fidelity 
at small perturbations.
However, all previous studies of the quantum system (\ref{eq3})
were done in the regime of quantum unitary evolution.
In contrast, for superconducting qubits and fluxonium, the dissipative
effects play a crucial role that leads us
to studies of quantum evolution (\ref{eq3})
in presence of dissipation present
for the kicked fluxonium.

The dissipative quantum evolution of oscillator systems
is well described in the frame of the Lindblad equation
for the density matrix $\rho(t)$ \cite{gorini,lindblad,weiss}.
In presence of dissipation with rate $\gamma$, a classical chaotic dynamics
in many cases converges to a strange attractor,
or chaotic attractor, with fractal structure on smaller and smaller scales
(see e.g. \cite{chaosbook,ott}).
The early studies of quantum strange attractor were
reported in \cite{graham1987,graham1988}
for the quantum Chirikov standard map with dissipation.
It was shown that a fractal stucture is washed
out on scales below Planck constant $\hbar$.
However, the properties of density matrix in this regime
were not investigated in detail.
The same model was studied in \cite{carlo}
in the frame of quantum trajectories \cite{carmichael,brun1996,brun2002}.

The results  in \cite{carlo},  obtained for dissipative quantum chaos, indicated
the existence of transition from Ehrenfest wave packet collapse to explosion.
In absence of dissipation the Ehrenfest theorem \cite{ehrenfest} guaranties that
a compact wave packet follows a classical trajectory during a certain
Ehrenfest time $t_E$. However, for systems with dynamical chaos
classical trajectories diverge rapidly due to exponential instability of motion
so that the Ehrenfest time is logarithmically short
$t_E \sim |\ln \hbar|/\Lambda$ comparing to a case of integrable dynamics where
$t_E \propto 1/\hbar$ is polynomially large at small values of Planck constant
(see e.g. \cite{chirikov1988,dlsehrenfest}). Here $\Lambda$ is the Lyapunov exponent
which characterizes the exponential instability of classical chaotic dynamics.
For unitary time evolution   in the regime of quantum chaos (at $\gamma=0$)
the illustrations of the Ehrenfest explosion of quantum wave packet
can be find e.g. at \cite{dlsehrenfest,husimi2,husimi3}.

In presence of dissipation and quantum chaos the results obtained with
quantum trajectories description show that the wave packet is collapsed
when the dissipative time $t_\gamma =1/\gamma$ is shorter than the Ehrenfest time $t_E$ \cite{carlo}:
\begin{align}
  t_\gamma & = 1/\gamma < t_E \sim |\ln \hbar|/\Lambda \;\; (collapse)\\
  t_\gamma &= 1/\gamma > t_E \sim |\ln \hbar|/\Lambda \;\; (explosion)
\label{eqcollapse}
\end{align}
This result was obtained with quantum trajectories
and it is important to understand its manifestation in the frame of
density matrix described by the Lindblad equation providing a complete
description of dissipative quantum evolution
which, as we show, converges to a quantum strange attractor for
the dissipative quantum system based on the Hamiltonian (\ref{eq3}).
In this work we describe the properties of the density matrix in this regime.
We also argue that the quantum strange attractor of this system
can be realized with kicked fluxonium or ion traps.
Here we consider only the case with $R=4$.

The article is organized as follows: Section II describes the model
and numerical computation methods of the Lindblad evolution,
results are presented in Section III and
discussion id given in Section IV.

\section{Model description} 
\label{sec2}

For classical dynamics the time evolution between
kicks is described by the equations of dissipative harmonic oscillator:
\begin{equation}
  d p/dt + 2\gamma p + {\omega_0}^2 x = 0 \; , \; d x/dt =p 
\label{eqcldis}
\end{equation}
where $\omega_0 =1$ is the frequency of free oscillator in (\ref{eq3}).
The equations are linear and their exact solution \cite{landaumech} gives an exact map of variable values
$(x,p)$ at the beginning of time between kicks $T$
to their values $\tilde{x}, \tilde{p}$ after period $T = 2\pi/R = \pi/2$ which
reads:
\begin{align}
\label{eqclmap1}
  {\tilde x} & = a \exp(-\pi\gamma/2) \cos(\pi\omega/2 + \alpha) ,\\
    {\tilde p} & = a \omega \exp(-\pi\gamma/2) \sin(\pi\omega/2 + \alpha) ,\\
    a  & = \sqrt{x^2 + (p + \gamma x)^2/\omega^2} ' \\
    \tan \alpha & = - (p+ \gamma x)/(\omega x) \; , \; \omega =\sqrt{1 - \gamma^2} \; .
\end{align}
In absence of dissipation the free dynamics simply rotate $(x,p)$ values on an angle $\pi/2$
on a circle in the phase plane. The kick after free propagation on time $T$ gives
the final value $(\bar{x}, \bar{p})$ after one period of full evolution:
\begin{align}
 \bar{p} = {\tilde p} - K q\sin{q\tilde{x}} \; , \; \bar{x} = \tilde{x} \; .
\label{eqclmap2}
\end{align}
Thus equations (\ref{eqclmap1})-(\ref{eqclmap2}) describe the full classical dynamics
on one time period with free propagation and kick. Iteration of these equations gives dynamical evolution
on many periods measured by integer time $t/T$ given by the number of kicks.

In presence of dissipation the quantum evolution of Hamiltonian ${\hat{H_f}}$ (\ref{eq3})
is described by the Lindblad equation for the density matrix $\rho$:
\begin{align}
\frac{\partial {\hat \rho}}{\partial{t}} &= -{\frac{i}{\hbar}}[ {\hat H}, {\hat \rho} ] 
+  2 \gamma \left({\hat a} {\hat \rho} {\hat a}^+ 
-  {\hat a}^+ {\hat a} {\hat \rho}/2 
-  {\hat \rho}  {\hat a}^+  {\hat a}/2 \right)
 \label{eqlindblad1}
\end{align}
where ${\hat a}, {\hat a}^+$ are oscillator operators and
$\gamma$ is the dissipation rate corresponding to those of
classical dissipative dynamics (\ref{eqcldis}).
During the propagation between kicks we have in the oscillator eigenbasis:
\begin{align}
&\frac{\partial \rho_{nm}}{\partial{t}} = i  (m - n)\rho_{nm} \nonumber \\
  \noindent
  &+ 2 \gamma \left(\sqrt{(n+1)(m+1)} \rho_{n+1,m+1} - (n + m) \rho_{n,m}/2 \right)
 \label{eqlindblad2}
\end{align}

Rewriting this equation in the interaction representation frame we have:
\begin{align}
  &\rho(t) = {\tilde \rho}(t) e^{i \omega_0  (m - n) t} \nonumber \\
  &\frac{\partial {\tilde \rho}_{nm}}{\partial{t}} = 2 \gamma \left(\sqrt{(n+1)(m+1)}
  {\tilde \rho}_{n+1,m+1} - (n + m) {\tilde \rho}_{n,m}/2 \right)
 \label{eqlindblad3}
\end{align}
An efficient integration of propagator is done in each 
independent $k$ indexed sub-block ${\tilde \rho}_{n, n+k}$ ($n$ being an integer in the basis).

During the kick the density matrix is changes to:
\begin{align}
  {\hat \rho} \rightarrow \exp\left( i K/\hbar \cos  q{\hat x}  \right) {\hat \rho}
  \exp\left( -i K/\hbar \cos  {q\hat x}  \right)
   \label{eqlindblad4}
\end{align}

The effect of kick is computed in the  oscillator eigenbasis
$\psi_n(x) =  e^{-x^2/2\hbar} H_n(x/\sqrt{\hbar})/[\pi^{1/4} \sqrt{2^n n!}] $
using the matrix elements (see \cite{prudnikov}):
\begin{align}
&\int_{-\infty}^{\infty} \cos q x \; \psi_n(x) \psi_{n+m}(x) d x \nonumber \\
&=\frac{1 + (-1)^m}{2} 2^{-m/2} \sqrt{\frac{n!}{(m+n)!}} \hbar^{m/2} q^m e^{-\hbar q^2/4} L_n^{m}(\hbar q^2/2)
   \label{eqlindblad5}
\end{align}
where $H_n$ and $L_n^{m}$ are Hermite  and  Laguerre polynomials respectively.

\begin{figure}[t]
\begin{center}
  \includegraphics[width=0.48\textwidth]{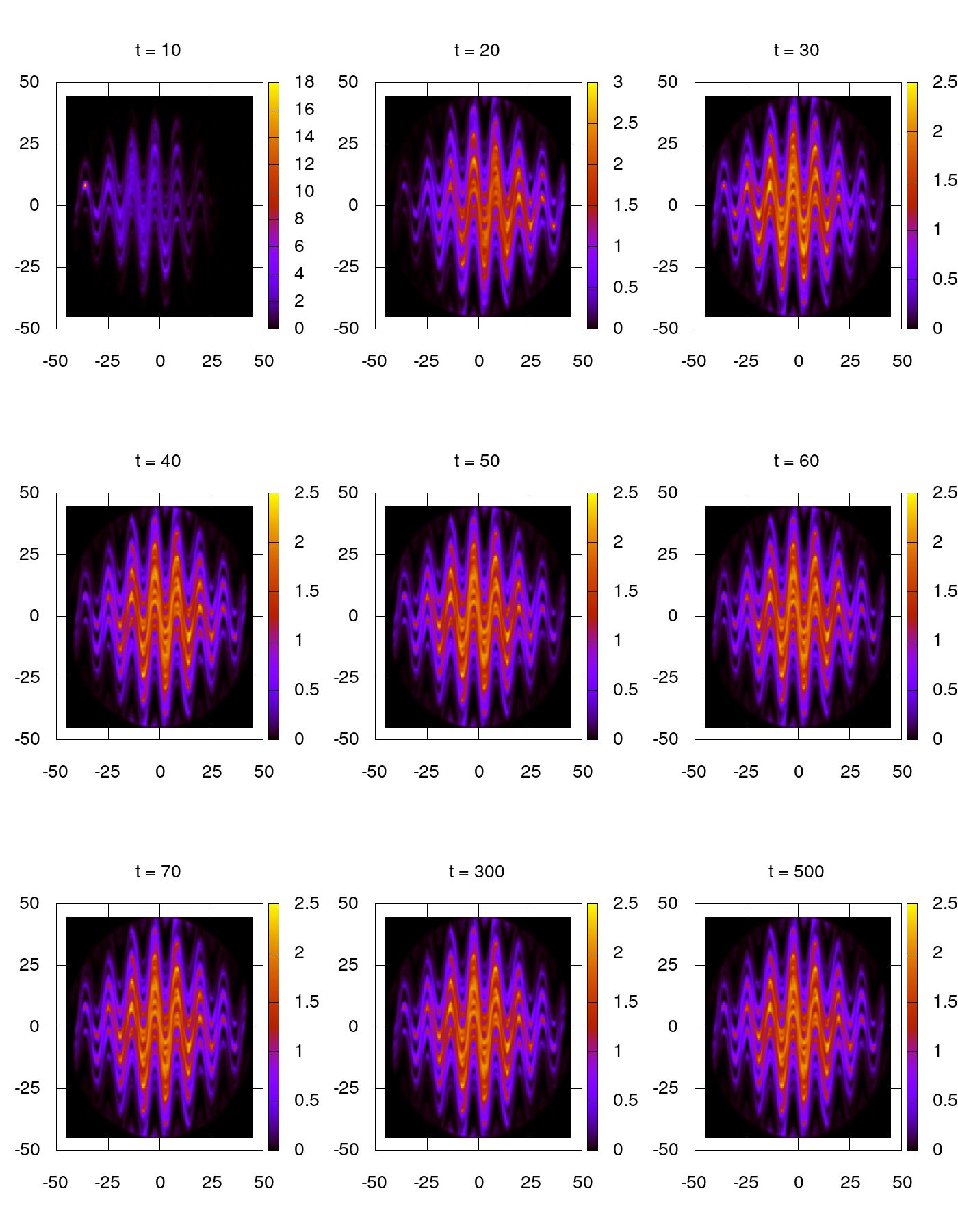}
\end{center}
\vglue -0.3cm
\caption{\label{fig1}
  Time evolution of Husimi function (shown at different time moments $t$)
  in the phase plane $(x,p)$,  $\hbar=\omega=1$,
   $k=K/\hbar =40$, $q=0.4$, $\gamma=0.05$,  $N=2000$
  ($\hbar_{eff} = \hbar q^2$, thus the classical chaos parameter
  rescaled to the case $q=1$ is $K_{cl}=K q^2 = 6.4$);
  here $t$ gives a number of kicks; $R=4$. Initial state at $t=0$ is a
  minimal coherent state located at $x=10, p=1$; color bars show Husimi function multiplied by
  factor $10^3$.
}
\end{figure}

We adapted the numerical integration code developed in \cite{qsync}
to perform numerical simulations of the Lindblad equation for
kicked fluxonium. The numerical effort scales as $O(N)$ 
where $N$ is the total number of oscillator eigenstates
while the total number of density matrix $\rho$ components
scales as $N_L = N^2$. We used up to $N=2000$ oscillator
eigenstates in our numerical simulations
that with up to $N_L=4 \times 10^6$ components of the density matrix.

By construction from Eqs.~(\ref{eqlindblad2})-(\ref{eqlindblad5})
the density matrix operator ${\tilde \rho}$, or density matrix,
is Hermitian. Thus due to the standard norm of $\rho$
its eigenvalues $\lambda_i$ are real being in the range
$0 \leq \lambda_i \leq 1$ with the trace
$Tr[{\tilde \rho}] = \sum_i \lambda_i =1$.   

As an initial quantum state we usually take a coherent oscillator state
with minimal size located at a certain $x,p$ position.
Classical and quantum evolution are converging to the same
steady-state strange attractor. The classical density distribution
in the phase space is obtained with $4 \times 10^6$ trajectories.

\section{Results} 
\label{sec3}

To characterize quantum time evolution we construct from the density matrix
$\rho(t)$ at time moment $t$
the Husimi function, which is obtained from the Wigner function
by a smoothing over $\hbar$ scale as described e.g. in \cite{husimi1,husimi2}.
Another more direct way is the trace of density matrix $\rho$
with a minimal coherent state $|\alpha><\alpha|$
located at various positions $\alpha=x+ip$.
The time evolution of the Husimi function is shown in Fig.~\ref{fig1}
for the regime of dissipative quantum chaos at classical chaos parameter
$K=6.4$ and dissipation $\gamma=0.05$. The data shows that the steady-state
distribution is reached approximately after $40$ kicks.

\begin{figure}[t]
\begin{center}
  \includegraphics[width=0.42\textwidth]{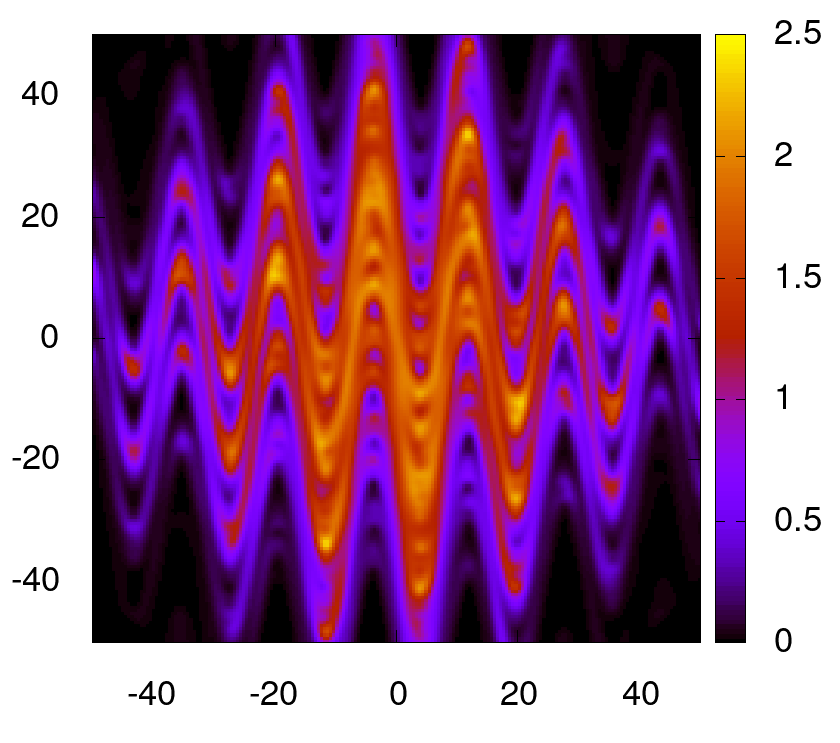}\\
\hskip 0.3cm   \includegraphics[width=0.44\textwidth]{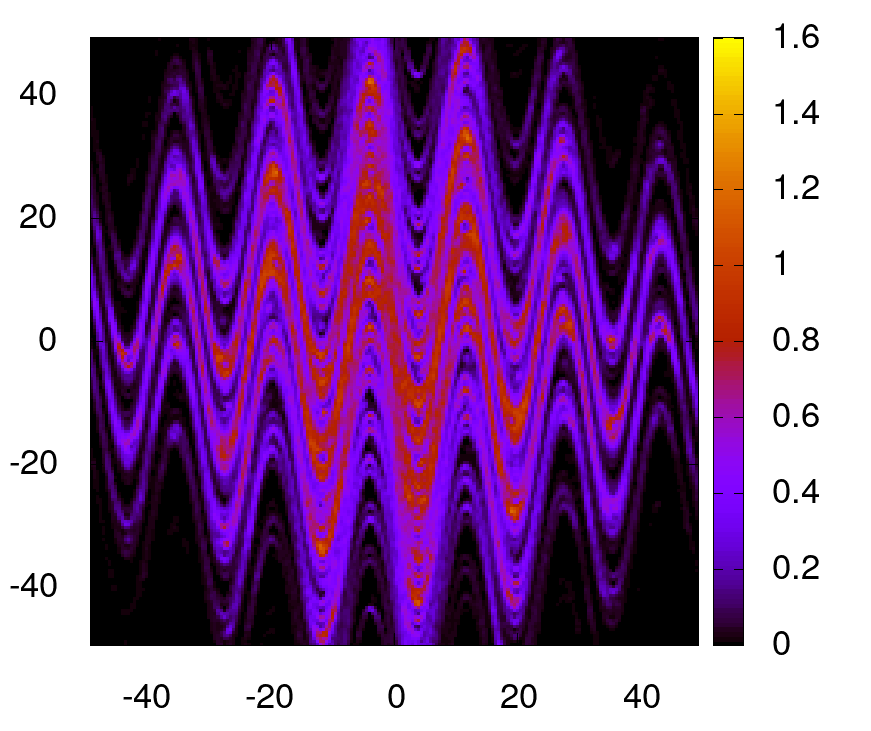}
\end{center}
\vglue -0.3cm
\caption{\label{fig2}
  Top panel: quantum Husimi function at the steady-state at $t=10^3$
  with parameters and notations of Fig.~\ref{fig1};
  bottom panel: classical density distribution 
  obtained with $4\times 10^6$ trajectories;
  all parameters are as in Fig.~\ref{fig1}.
  Color density values are increased by a factor $10^3$.
}
\end{figure}

The quantum steady-state of Husimi function is shown at
large time $t=1000$ at the top panel of Fig.~\ref{fig2}.
The corresponding classical distribution in the phase space $(x,p)$
is shown at the bottom panel of Fig.~\ref{fig2}.
The comparison of two cases shows that the dissipative quantum
distribution is very close the classical one.

The classical case corresponds to a strange chaotic attractor
for which the fractal information dimension
can be estimated as
$d_1= 2 - \gamma/\Lambda$ \cite{chaosbook,ott}.
Here the Lyapunov exponent is
approximately $\Lambda \approx \ln(K_{cl}/2) \approx 1$
(as for the Chirikov standard map \cite{chirikov1979}).
Thus we have $d_1 \approx 1.95$ for the case of Fig.~\ref{fig2}
at $K_{cl}=6.4$, $\gamma=0.05$.
The diffusion growth of system energy $E \approx <p^2> \sim q^2 K^2t/2$
is stopped by dissipation at time $t_\gamma \sim 1/\gamma$
that gives the distribution width in momentum $p$ (and coordinate $x$)
being $\Delta p \sim q K /\sqrt{2\gamma} \sim 50$
and being close to the distribution width $ p \sim \pm 50$
obtained numerically in Fig.~\ref{fig2} at corresponding parameters.
We do not present detailed discussion of classical dissipative dynamics
since the properties of strange chaotic attractors
had been studied deeply for many system as reported in
\cite{chaosbook,ott} and since our main aim is the analysis
of quantum properties of density matrix evolution
given by the Lindblad equations (\ref{eqlindblad1})-(\ref{eqlindblad4}).

\begin{figure}[t]
\begin{center}
  \includegraphics[width=0.48\textwidth]{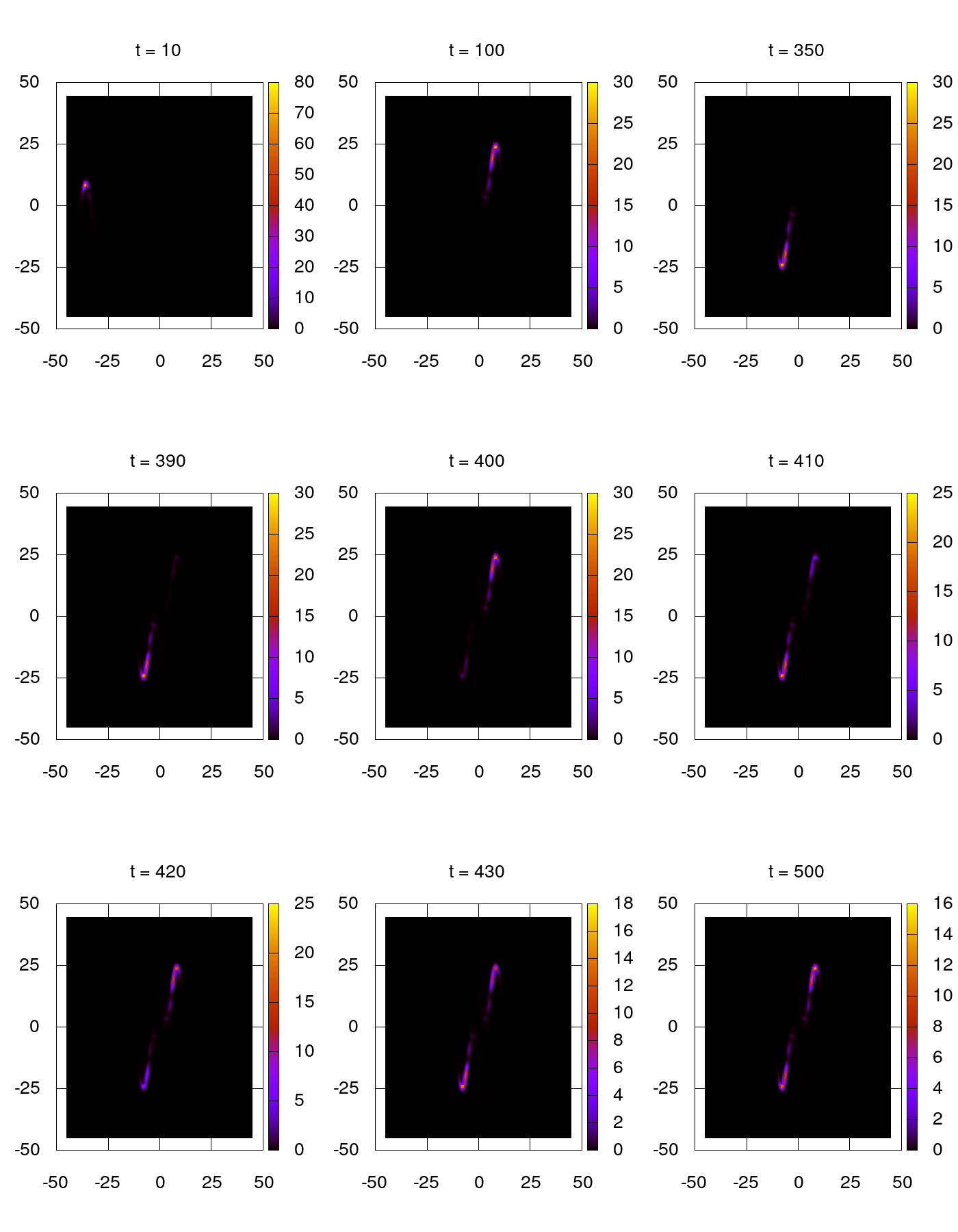}
\end{center}
\vglue -0.3cm
\caption{\label{fig3}
  Collapse of density matrix eigenstate at maximal eigenvalue $\lambda_i$;
  Husimi function of eigenstate is shown at time moments $t$,
  color shows its values increased by a factor $\times 10^3$,
  system parameters are as in Fig.~\ref{fig1}.
}
\end{figure}

To analyze the properties of density matrix ${\hat \rho(t)}$ we 
compute its eigenvectors with eigenvalues $0 \le \lambda_i \le 1$
at time moments $t$
(${\hat \rho(t)} \chi_i (t) = \lambda_i(t) \chi_i(t))$.
A typical example of a Husimi function time evolution of eigenvector $\chi_i$
at maximal $\lambda_i$ is shown in Fig.~\ref{fig3}.
The main feature of such an eigenstate of $\rho(t)$
is its localization, or collapse,  in the phase space $(x,p)$.
Other eigenstates also have a similar localized structure.
With time the localized wave packet splits
on two packets located at symmetric positions
of maximum at $(x_m,p_m)$ and  $(-x_m,-p_m)$.
This can be viewed as a formation of Schr\"odinger cat
eigenstates of density matrix 
(see results for cat states
e.g. in \cite{mazyar} and Refs. therein).
This symmetry corresponds to a symmetry of system Hamiltonian (\ref{eq3}).
The steady-state distribution of $\chi_i$ is formed
at relatively large times $t_{cat} \approx 420$.
This time is significantly large than the relaxation time
$t_\gamma \sim 1/\gamma \sim 20$ at which the global steady-state
is reached for ${\hat \rho(t)}$ in Fig.~\ref{fig1}.

\begin{figure}[t]
\begin{center}
  \includegraphics[width=0.48\textwidth]{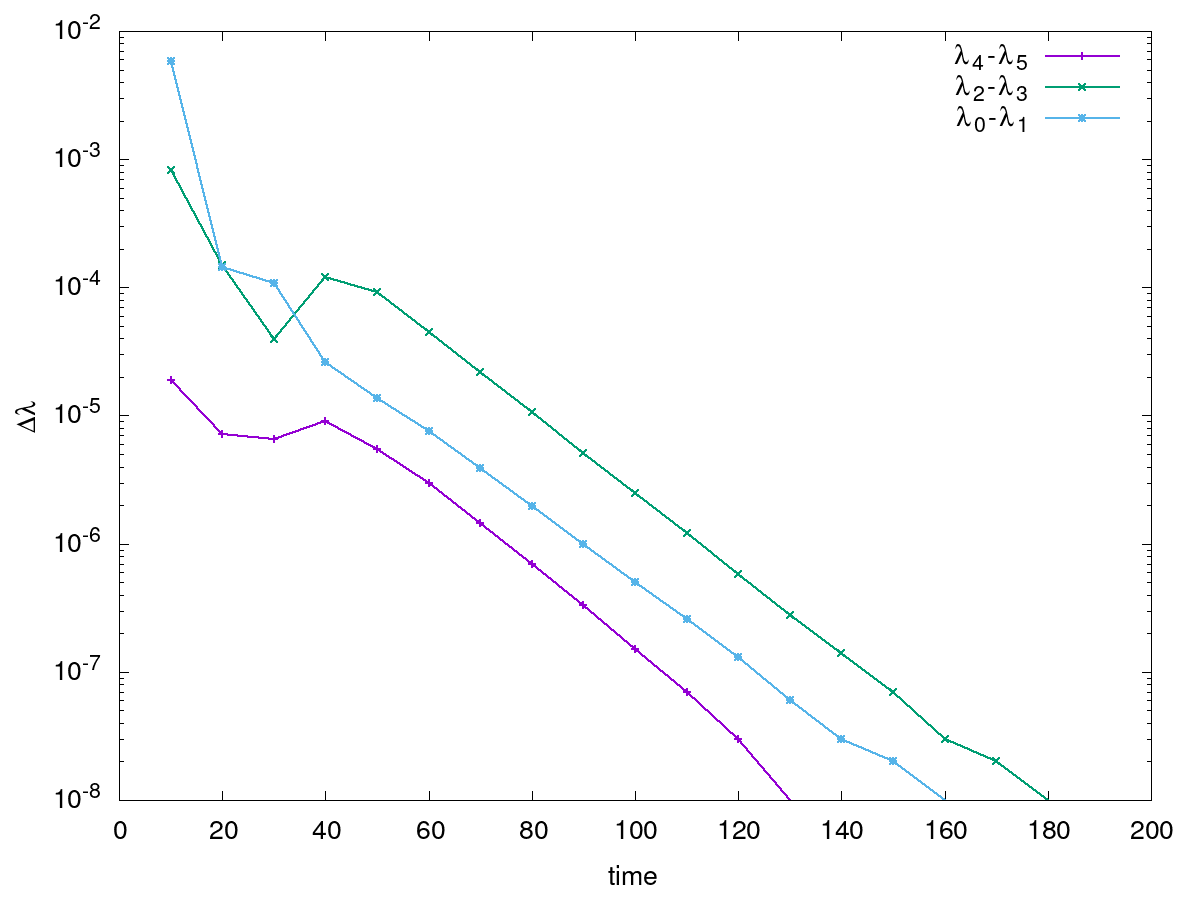}
\end{center}
\vglue -0.3cm
\caption{\label{fig4}
  Splitting $\Delta \lambda$ of largest eigenvalues  $\lambda_i$ of density matrix $\rho(t)$
  at different moments of time $t$ (given here  in number of kicks); three quasi-degenerate pairs
  of largest eigenvalues are shown: $\lambda_0 -\lambda_1$,   $\lambda_2 -\lambda_3$,
  $\lambda_4 -\lambda_5$; system parameters are as in Fig.~\ref{fig1}.
}
\end{figure}

The reason of large value of $t_{cat} \gg t_\gamma$
becomes clear from the results presented in Fig.~\ref{fig4}.
Indeed, this data shows that with time there
appears a strong quasi-degeneracy between
the pairs of eigenvalues of density matrix.
Thus  initial asymmetric eigenstates pf ${\hat \rho(t)}$
relax to symmetric ones at large times $t$.
This relaxation is slow due to quasi-degeneracy
of cat eigenstates.
The symmetry of eigenstates correspond
to the Hamiltonian symmetry $x \rightarrow -x$
preserved in presence of dissipation.

\begin{figure}[t]
\begin{center}
  \includegraphics[width=0.42\textwidth]{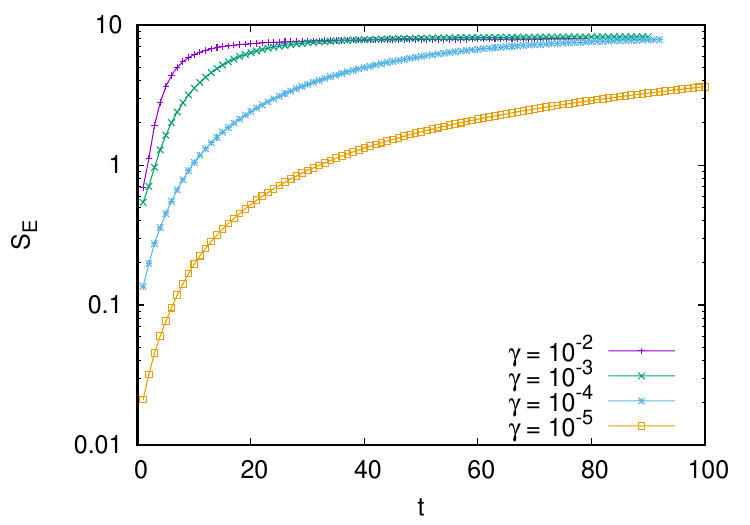}
\end{center}
\vglue -0.3cm
\caption{\label{fig5}
  Entropy of entanglement $S_E$ vs time $t$,
  for system parameters $K=8$, $\hbar=q=1$, values of $\gamma$ are given in the panel.
}
\end{figure}

We also characterize the density matrix ${\hat \rho(t)}$
by its entropy of entanglement given by
$S_E(t) =  -Tr[{\hat \rho(t)} \ln {\hat \rho(t)}] =
- \sum_i \lambda_i \ln \lambda_i$ \cite{nielsen,qwiki1}.
The dependence of $S_E(t)$ on time $t$ is shown
in Fig.\ref{fig5}. Here the initial state is unitary
and the breaking of unitarity is slow at small $\gamma$.
Due to this at small times $S_E$ is smaller at smaller $\gamma$
values, but at large times due to chaos $\rho(t)$
spreads over all available system basis at $N=2000$
reaching maximal possible $S_E$ values.

However, it should be noted that large values of $S_E$
do not imply that the system is really quantum.
The right characteristic is the quantum negativity $G_N$ \cite{nielsen,qwiki2}
computed via a partial transpose $T_g$ of density matrix with respect to a subsystem.
We can virtually introduce a spin half for the system described by the Lindblad operator $\cal{L}$
which we notes as $|\alpha>, |\beta>$.
Then we perform certain transformations that lead to the expression for $G_N$:

\onecolumngrid

\begin{align}
  \frac{|g_0  \alpha \rangle + |g_1  \beta \rangle}{\sqrt{2}}  &\xrightarrow[]{\rho}
  \frac{1}{2} \left( |g_0\rangle \langle g_0| \otimes |\alpha\rangle \langle \alpha| + 
  |g_0\rangle \langle g_1| \otimes |\alpha\rangle \langle \beta| + |g_1\rangle \langle g_0| \otimes |\beta\rangle \langle \alpha| + \nonumber 
  |g_1\rangle \langle g_1| \otimes |\beta\rangle \langle \beta| \right) \nonumber \\
  & \xrightarrow[]{\cal{L}}  \frac{1}{2} \left\{ |g_0\rangle \langle g_0| {\cal L}(|\alpha\rangle \langle \alpha|) +  |g_0\rangle \langle g_1| {\cal L}(|\alpha\rangle \langle \beta|) + |g_1\rangle \langle g_0| {\cal L}(|\beta\rangle \langle \alpha|) + |g_1\rangle \langle g_1| \cal{L}(|\beta\rangle \langle \beta|) \right\} \nonumber \\
  & \xrightarrow[]{T_g} \rho^{T_g} \defeq \frac{1}{2} \left\{ |g_0\rangle \langle g_0| {\cal L}(|\alpha\rangle \langle \alpha|)^{t} +  |g_0\rangle \langle g_1| {\cal L}(|\alpha\rangle \langle \beta|)^{t} + |g_1\rangle \langle g_0| {\cal L}(|\beta\rangle \langle \alpha|)^{t} + |g_1\rangle \langle g_1| {\cal L}(|\beta\rangle \langle \beta|)^{t} \right\} \nonumber \\
  { G_N}(t) &= \left|\sum_{\lambda_{n(\rho^{T_g}) < 0} } \lambda_n(\rho^{T_g})\right| 
\end{align}

\twocolumngrid

\begin{figure}[t]
\begin{center}
  \includegraphics[width=0.42\textwidth]{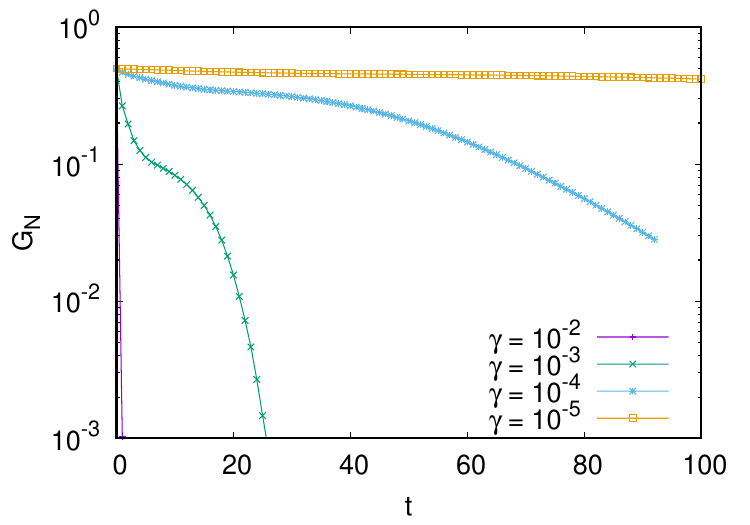}
\end{center}
\vglue -0.3cm
\caption{\label{fig6} Dependence of quantum negativity $G_N$ on time $t$
  for system parameters $K=8$, $\hbar=q=1$, values of $\gamma$ are given in the panel. 
}
\end{figure}

The  dependence $G_N(t)$ on time $t$ is presented in Fig.~\ref{fig6}.
The results show that quantum negativity drops very rapidly with time going to almost
zero for $\gamma=0.01, 10^{-3}$. For smaller values of $\gamma=10^{-4}, 10^{-5}$
$G_N(t)$ has finite values
since the time scale shown in Fig.~\ref{fig6}
is small compared to the dissipative time $t_\gamma =1/\gamma$.
These results show that the quantum negativity becomes almost zero
for times $t > t_\gamma$. Thus the quantum features of evolution
are washed out for times $t> t_\gamma$. On such scales
the quantum Lindblad evolution
is similar to a classical wave packet evolution
in presence of dissipation and classical noise
which amplitude corresponds to
amplitude of quantum dissipative fluctuations.

\section{Discussion} 
\label{sec4}
Here we presented studies of properties of density matrix in the regime
of dissipative quantum chaos. We find that
at strong or moderate dissipation the density matrix in the steady-state
regime describes a quantum strange attractor.
In the phase space above the scale of Planck constant 
its structure well reproduces those of the classical strange attractor.
We show that in this regime the eigenstates of density matrix
are localized in the phase space. This localization is argued to
reflect the quantum wave packet localization
obtained in the frame of quantum trajectories
discussed in \cite{carlo}. 
It is found that in this regime the entropy of entanglement $S_E$
grows with time reaching its maximal value related to a size
of strange attractor in the phase space.
At the same time the quantum negativity $G_N$
drops rapidly with time to zero in this regime.
Due to numerical restrictions we do not present the
regime of weak dissipation where we expect to have
the Ehrenfest explosion or delocalization of eigenstates of
density matrix since this regime requires
long integration times and large numerical basis.
At the same time for unitary evolution
at $\gamma=0$ the Ehrenfest explosion
of wave packets is well established (see e.g. \cite{dlsehrenfest,husimi2,husimi3})
and we expect that at weak dissipation for quantum chaos
the eigenstates of density matrix become delocalized.

Even if the density matrix eigenstates are localized
at strong or moderate dissipation
the whole density matrix well reproduces
the structure of the classical strange attractor.
Specific experimental methods should be developed
to detect the localized structure of density
matrix eigenstates in this strange attractor regime.
We hope that a significant experimental progress
with the fluxonium studies
\cite{lyon,nguyen,alibaba,quentin2023,vavilov2024,gatemonium}
  will allow to investigate experimentally
  quantum strange attractor of fluxonium.

\noindent {\bf Acknowledgments:}
The authors acknowledge support from the grants
 ANR France project OCTAVES (ANR-21-CE47-0007),
NANOX $N^\circ$ ANR-17-EURE-0009 in the framework of 
the Programme Investissements d'Avenir (project MTDINA),
MARS (ANR-20-CE92-0041) and EXHYP (INP Emergence 2022).


\end{document}